\documentclass[aps,
prd,
12pt,
final,
oneside,
onecolumn,
nobibnotes,
nofootinbib,
superscriptaddress,
centertags,
floatfix,
secnumarabic,
notitlepage,
showkeys]{revtex4-2}%

\usepackage{graphicx}
\usepackage{dcolumn}
\usepackage{bm}
\usepackage{amsmath}
\usepackage{xcolor}
\usepackage{hyperref}
\usepackage{comment}

\begin{document}

\title{Production of dileptonic bound states in the Higgs boson decay}

\author{F. A. Martynenko}%
\email{f.a.martynenko@gmail.com}
\affiliation{Samara University}
\author{A. P. Martynenko}
\email{a.p.martynenko@samsu.ru}
\affiliation{Samara University}
\author{A. V. Eskin}%
\email{eskinalexey1992@gmail.com}
\affiliation{Samara University}



\date{\today}

\begin{abstract}
The production of single and paired lepton bound states in the decay of the Higgs boson has been studied. We explore different decay mechanisms that contribute significantly to the decay width. The decay widths are calculated taking into account relativistic corrections in the decay amplitude and in the wave function of the bound state of leptons.
\end{abstract}

\maketitle


\section{Introduction}
\label{intro}

After the discovery of the Higgs boson, a period of more detailed study of its properties and decay characteristics through various channels began. Such theoretical research now relies on existing experimental data, the collection of which is steadily expanding \cite{higgs}. Plans have begun to emerge for various future accelerators in which there will be a significant increase in the production of Higgs bosons, and, therefore, the possibility of more accurately determining various parameters of the Higgs sector will open up \cite{clic,fcc}. As often happens when formulating programs for future colliders, the task is set to search for events that do not fit into the scheme of the Standard Model. Among the rare decays of the Higgs boson, one can distinguish a class of processes in which bound states of fundamental particles arise, such as mesons and baryons \cite{p1,p2,p3,p4}. The theoretical interest in these reactions is due to the fact that, among other things, they can additionally test models for describing bound states of quarks. The number of such works has grown significantly over the past 10 years, and the methods for calculating the observed characteristics have received significant development \cite{higgs1}. 
Another group of reactions for the formation of bound states of particles consists of those reactions in which bound states of leptons (positronium, dimuonium, ditauonium) can be produced. In reactions of this type there are no uncertainties caused by the nonperturbative interaction of quarks.
Various processes of production and decay of positronium already have a rich history, which included critical periods of divergence between theory and experiment. Other bound states of leptons have been studied much less. Thus, it can be noted that calculations of the characteristics of dimuonium have been carried out quite a long time ago with great accuracy, but the bound state itself has not yet been observed \cite{sgk1}. Various concepts of dimuonium and ditauonium production are discussed in many papers \cite{higgs2,higgs3,enterria}.
One possibility for the production of lepton bound states is connected with the decay of the Higgs boson.
In this work, we study the processes of both single and pair production of dimuonium and ditauonium within the framework of relativistic method of describing the production of bound states of particles. This method was used previously to calculate the production of charmonium and $B_c$ mesons \cite{apm2021,apm2023,apm2023a}. It allows one to obtain values of the decay width taking into account relativistic effects. It is useful to note that the observation of orthodimuonium or orthoditauonium is connected with subsequent decay into three photons.

\section{Production of single leptonic bound states}

Let us first consider the single production of lepton bound states in the decay of $H \to \gamma+\bar l l$, $H \to Z+\bar l l$. The Higgs boson-lepton vertex is determined by the factor \cite{roma}:
\begin{equation}
m(\sqrt{2}G_F)^{1/2}=\frac{e}{\sin 2\theta_W}\frac{m}{M_Z},
\end{equation}
where $m$ is the lepton mass, $M_Z$ is the Z-boson mass,
$\theta_W$ is the Weinberg angle, $G_F$ is the 
Fermi coupling constant.

\begin{figure}[htbp]
\includegraphics[width=0.97\textwidth]{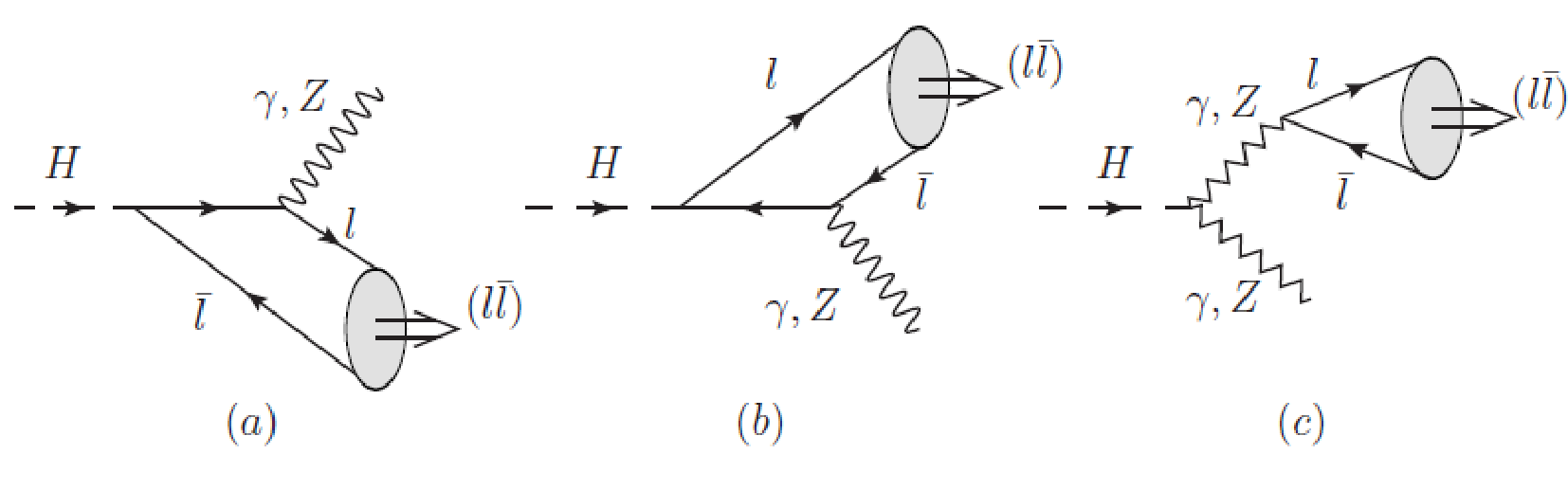}
\caption{Different mechanisms of single leptonium production in the Higgs boson decay.}
\label{fig1}
\end{figure}

The lepton-photon (lepton-Z-boson) mechanism of single production of dimuonium and ditauonium is determined by three amplitudes presented in Fig.~\ref{fig1}(a,b,c). The initial expression for the amplitude of such a process can be represented as a convolution of the amplitude of the production of a photon and a lepton-antilepton pair with the Bethe-Salpeter wave function, which describes, within the framework of the four-dimensional formalism, a lepton-antilepton bound state moving with 4-momentum $P$. 
Transforming the amplitude during the transition to three-dimensional formalism
and using the law of transformation of the quasipotential wave function during the transition from the moving frame of reference to the rest frame of the lepton bound state (see, \cite{faustov,brodsky}), we obtain two decay amplitudes (Fig.~\ref{fig1}(a,b)) in the form:
\begin{gather}
\label{amp1}
{\cal M}_1(k,P)=\frac{4\pi\alpha m_l}{\sin 2\theta_W M_Z}
\int\frac{d{\bf p}}{(2\pi)^3}\frac{\Psi({\bf p})}{\frac{\varepsilon}{m}\frac{(\varepsilon+m)}{2m}[(r-p_2)^2-m^2]}\times \\
Tr\bigl\{
[\frac{\hat v-1}{2}-\hat v\frac{p^2}{2m(\varepsilon+m)}-
\frac{\hat p}{2m}]\hat\varepsilon_{l\bar l}\frac{(\hat v+1)}{2\sqrt{2}}
[\frac{\hat v+1}{2}- 
\hat v\frac{p^2}{2m(\varepsilon+m)}+\frac{\hat p}{2m}]\hat\varepsilon_\gamma(\hat r-\hat p_2+m)\bigr\}  \nonumber, 
\end{gather}
\begin{gather}
\label{amp2}
{\cal M}_2(k,P)=\frac{4\pi\alpha m_l}{\sin 2\theta_W M_Z}
\int\frac{d{\bf p}}{(2\pi)^3}\frac{\Psi({\bf p})}{\frac{\varepsilon}{m}\frac{(\varepsilon+m)}{2m}[(p_1-r)^2-m^2]}\times \\
Tr\bigl\{
[\frac{\hat v-1}{2}-\hat v\frac{p^2}{2m(\varepsilon+m)}-
\frac{\hat p}{2m}]\hat\varepsilon_{l\bar l}\frac{(\hat v+1)}{2\sqrt{2}}
[\frac{\hat v+1}{2}-
\hat v\frac{p^2}{2m(\varepsilon+m)}+\frac{\hat p}{2m}](\hat p_1-\hat r+m)\hat\varepsilon_\gamma\bigr\} \nonumber, 
\end{gather}
where $r$ is the Higgs boson four-momentum, $\varepsilon_\gamma$ is the photon polarization vector, $\varepsilon_{l\bar l}$ is the polarization vector of dimuonium (ditauonium), $k$ is the photon four-momentum, $\varepsilon(p)=\sqrt{p^2+m^2}$, $v=P/M$, $M$ is the mass of $l\bar l$ state. $p$ is the lepton relative four-momentum. The expressions for the amplitudes \eqref{amp1}, \eqref{amp2}
are obtained as a result of the Lorentz transformations in the same way as in our works \cite{apm2021,apm2023,apm2023a}.
They make it possible to take into account relativistic corrections connected with the momentum of relative motion of leptons $p$.
The 4-momenta of the photon and leptonium (bound state lepton-anti-lepton) are given as arguments in the amplitudes \eqref{amp1}, \eqref{amp2}, although it should be remembered that they lie on the mass surface, and the following conservation law is satisfied: $r=k+P$.

The simplifications that can be made in the amplitudes are primarily related to the denominators of the lepton propagators. The magnitude of the relative momentum is small compared to the mass of the Higgs boson $M_H$, so we have:
\begin{equation}
\label{appr}
(p_1-r)^2-m^2\approx (r-p_2)^2-m^2 \approx \frac{1}{2}(M_H^2-M^2),
\end{equation}
where the lepton mass can be approximately replaced here by $m=\frac{M}{2}$ neglecting bound state effects. 
We completely neglect corrections of the form $|{\bf p}|/M_H$. At the same time, we keep further in the decay amplitudes the second-order correction 
for small ratios $|{\bf p}|/m$ relative to the leading order result.
If it is necessary to increase the accuracy of the calculation, the corrections $|{\bf p}|/M_H$ can also be taken into account within the used method.

After calculating the trace and extracting relativistic corrections in $p/m$ of second and fourth order in the numerator, we obtain the following expression for the total decay amplitude:
\begin{gather}
{\cal M}=\frac{64\pi\alpha r_1}{\sin 2\theta_W M_Z(r_2^2-1)}\tilde\psi(0)\Biggl\{(\varepsilon_\gamma \varepsilon_{l\bar l})\Bigl[r_2^2\bigl(\frac{1}{2}-\frac{1}{6}\omega_1\bigr)-
r_1+r_1\omega_1+\frac{4}{3}r_1\omega_2\Bigr]-\nonumber \\
(v_\gamma\varepsilon)(v \varepsilon_\gamma)\Bigl[1-\frac{1}{3}\omega_1\Bigr]\Biggr\}, 
\label{samp1}
\end{gather}
where for convenience the particle mass ratios are introduced: $r_1=\frac{m}{M}$, $r_2=\frac{M_H}{M}$.
The expression \eqref{samp1} clearly demonstrates the general structure of the decay amplitudes, which contains two characteristic parts with $(v_\gamma\varepsilon)(v \varepsilon_\gamma)$ and $(\varepsilon_\gamma \varepsilon_{l\bar l})$.

Relativistic corrections in \eqref{samp1} of the second and fourth orders in $p/m$ are determined in our approach by the parameters $\omega_1$, $\omega_2$
and the value of the wave function at zero $\tilde\psi(0)$.
The definition of these parameters and their numerical calculation is given in the next section.
Note that we have retained the 4th order $\alpha$ corrections here only to demonstrate the overall amplitude structure. In what follows, when obtaining numerical estimates, we neglect fourth-order corrections.

Using the decay amplitude \eqref{samp1}, we can calculate the differential decay width according to the formula:
\begin{equation}
\label{dgamma}
d\Gamma=\frac{\vert{\bf P}\vert}{32\pi^2 M_H^2}\overline{|M|^2}d\Omega.
\end{equation}
Three-momentum of final leptonium state is expressed in terms of masses of the Higgs boson $(M_H)$, $Z$-boson $(M_Z)$, and $(l\bar l)$ state $(M)$:
\begin{equation}
\label{three}
\vert{\bf P}\vert=
\begin{cases}
    \sqrt{\frac{M^4+(M_H^2-M_Z^2)^2-2M^2(M_H^2+M_Z^2)}{4M_H^2}},~~~ H \rightarrow (l{\bar l}) + Z\\
    \frac{M_H^2-M^2}{2M_H},~~~ H \rightarrow (l{\bar l}) + \gamma \\
    \sqrt{\frac{M_H^2}{4}-M^2},~~~ H \rightarrow (l{\bar l}) + (l{\bar l})
\end{cases}
.
\end{equation}

In the numerator of the amplitude \eqref{samp1} we define two functions $g_{1,2}$ as follows:
\begin{gather}
\label{numer}
N=g_1(\varepsilon_{l\bar l}\varepsilon_\gamma)-g_2(v\varepsilon_\gamma)(v_\gamma\varepsilon_{l\bar l}), \\
g_1=\Bigl[r_2^2\bigl(\frac{1}{2}-\frac{1}{6}\omega_1\bigr)-
r_1+r_1\omega_1+\frac{4}{3}r_1\omega_2\Bigr],~~~
g_2=\Bigl[1-\frac{1}{3}\omega_1\Bigr]. \nonumber
\end{gather}

After summing over the polarization of the photon and ortho-dileptonium we obtain total decay width $H\to \gamma+(l\bar l)$ in the form:
\begin{gather}
\label{ga1}
\Gamma_{\gamma(l\bar l)}=\frac{512\pi\alpha^2 r_1^2}{r_2^3\sin^2 2\theta_W M_Z^2(r_2^2-1)}
\Biggl[
\frac{1}{2}r_2^2g_2^2+3g_1^2-\frac{1}{4}g_2^2-\frac{1}{4}r_2^4g_2^2
\Biggr]\vert\tilde\psi(0)\vert^2.
\end{gather}

The decay width of the Higgs boson into $Z$-boson and $(l\bar l)$ (see Fig.~\ref{fig1}(a,b)) is calculated similarly and has the following form:
\begin{gather}
\Gamma_{Z(l\bar l)}=\frac{32\pi\alpha^2
\sqrt{[(r_2+1)^2-r_3^2][(r_2-1)^2-r_3^2]}}{r_2^3
\sin^42\theta_W M_Z^2(r_2^2-1)^2}
\vert\tilde\psi(0)\vert^2
\Biggl\{
g_{2Z}^2(-\frac{ 1}{4}+\frac{1}{16}r_3^{-2}+
\frac{3}{8}r_3^2-\frac{1}{4}r_3^4+\nonumber  \\
\frac{1}{16}r_3^6-\frac{1}{4}r_2^2 r_3^{-2}+\frac{1}{4}r_2^2+ \frac{1}{4}r_2^2r_3^2-\frac{1}{4}r_2^2r_3^4+\frac{3}{8}r_2^4r_3^{-2}+\frac{1}{4}r_2^4+\frac{3}{8}r_2^4r_3^2-\frac{1}{4}r_2^6 r_3^{-2}-\frac{1}{4}r_2^6+
\frac{1}{16}r_2^8r_3^{-2})+\nonumber  \\
g_{1Z}g_{2Z}(-\frac{1}{4}+\frac{1}{4} r_3^{-2}-\frac{1}{4} r_3^2+\frac{1}{4} r_3^4\frac{3}{4} r_2^2 r_3^{-2}- \frac{1}{2} r_2^2 - 3/4 r_2^2 r_3^2 + \frac{3}{4}r_2^4r_3^-2+\frac{3}{4} r_2^4-\frac{1}{4} r_2^6r_3^{-2} )+\nonumber  \\
g_{1Z}^2( \frac{5}{2}+\frac{1}{4}r_3^-2+\frac{1}{4}r_3^2-\frac{1}{2} r_2^2r_3^{-2}-\frac{1}{2}r_2^2+\frac{1}{4}r_2^4r_3^{-2})
\Biggr\}, 
\label{ga2}
\end{gather}
\begin{gather}
g_{1Z}=(r_2^2-r_3^2)(\frac{1}{8}+\frac{1}{4}a_z-\frac{1}{24}\omega_1-\frac{1}{12}\omega_1 a_z)+
r_1(-\frac{1}{4}-\frac{1}{2}a_z+\frac{1}{4}\omega_1+\frac{1}{2}\omega_1 a_z),\nonumber \\
g_{2Z}=(\frac{1}{4}+\frac{1}{2}a_z-\frac{1}{12}\omega_1-\frac{1}{6}a_z\omega_1), 
\label{ga3}
\end{gather}
where we introduce another mass coefficient $r_3=\frac{M_Z}{M}$, $a_z=2\sin^2\theta_W$.

To obtain total width of the decay $H\to Z+(l\bar l)$, including the contribution of the $ZZ$-production mechanism (see, Fig.~\ref{fig1}(c)), it is necessary to make the following substitution in \eqref{ga2}:
\begin{gather}
g_{1Z}\to g_{1Z}+\frac{r_2^2}{r_1}g_{1ZZ},~~~
g_{1ZZ}=-\frac{1}{4}-\frac{1}{2}a_z-\frac{1}{12}\omega_1-
\frac{1}{6}\omega_1 a_z.
\label{ga4}
\end{gather}

The expressions \eqref{numer}, \eqref{ga1}-\eqref{ga3} clearly show that the main parameters that determine numerical values of decay branchings include, together with the fine structure constant, the particle mass ratios $r_1$, $r_2$, $r_3$. 

\begin{figure}[htbp]
\includegraphics[width=0.97\textwidth]{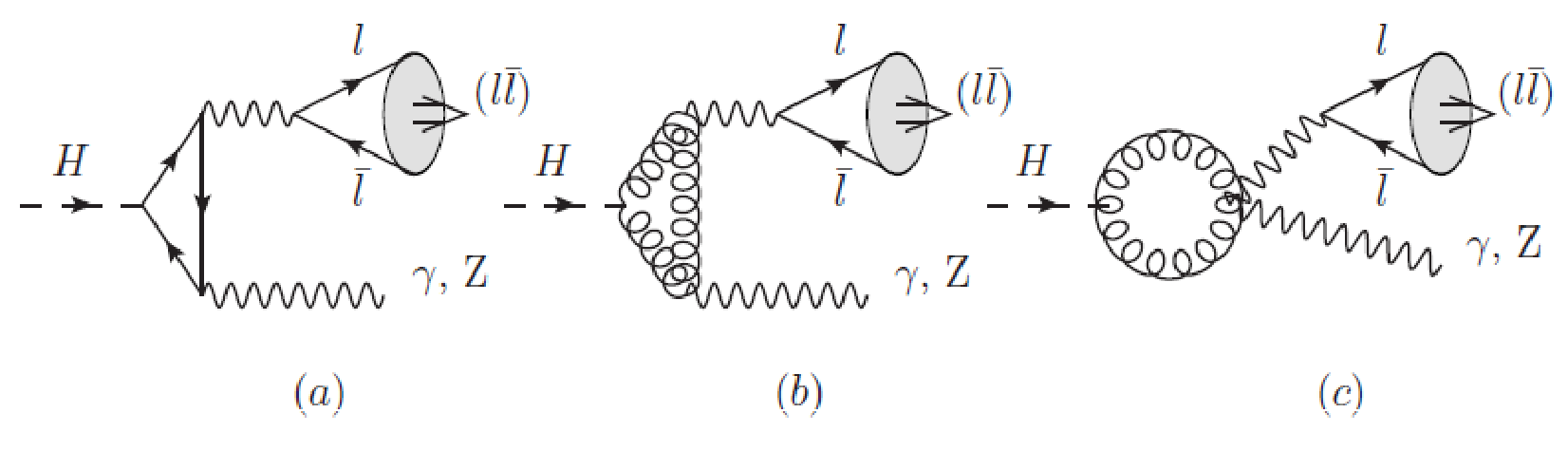}
\caption{Quark loop and $W$-boson loop mechanisms of single leptonium production in the Higgs boson decay.}
\label{fig11}
\end{figure}

The studies we carried out in \cite{apm2021,apm2023} show that it is necessary to take into account the loop ($W$-boson and quark) production mechanism 
$Z+(l\bar l)$ shown in Fig.~\ref{fig3}. Although formally an additional degree of $\alpha$ appears in it compared to the $ZZ$-boson mechanism, the fine structure constant is not a decisive factor in determining the order of magnitude of the contribution. It is useful to mention here that a loop mechanism with W-boson and quark loops is needed to describe the decay process $H\to Z+\gamma$, the first evidence for which was presented recently in \cite{Aad}.
Numerous particle mass ratio parameters begin to play a more significant role in loop production mechanism. The loops of heavy quarks and the $W$-boson make their contribution to the structure functions $g_1$ and $g_2$, which can be represented as follows:
\begin{equation}
\label{ga31}
g^{(loop)}_{1~Z}=\frac{\pi\alpha r_2^2 r_3^2\cos^2\theta_W}{2r_1}A^Z_W
(r_2^2-r_3^2)(1+\frac{1}{3}\omega_1)+\sum_{Q=c,b,t}
\frac{3\alpha q_Q r_2^2 m_Q^2}{r_1 M^2}A^Z_Q(r_2^2-r_3^2)(1+
\frac{1}{3}\omega_1),
\end{equation}
\begin{equation}
\label{ga31a}
g^{(loop)}_{2~Z}=\frac{\pi\alpha r_2^2 r_3^2\cos^2\theta_W}{r_1}A^Z_W
(1+\frac{1}{3}\omega_1)+\sum_{Q=c,b,t}
\frac{6\alpha q_Q r_2^2 m_Q^2}{r_1 M^2}A^Z_Q(1+\frac{1}{3}\omega_1), 
\end{equation}
where relative coefficients with respect to the lepton-Z-boson mechanism are indicated.
The determination of the loop parameters $A^Z_W$, $A^Z_Q$ is described below as in the case of the production of $\gamma+(l\bar l)$.
Numerical results of calculating the decay widths with the production of $Z+(l\bar l)$ are shown in  Table~\ref{tab_decay_width}.

The lepton-photon decay mechanism in Fig.~\ref{fig1} gives a decay width of fifth order in $\alpha$. But our analysis shows that the lepton-photon mechanism in Fig.~\ref{fig11} has a significantly larger numerical contribution, although it has a higher order $\alpha^7$. This occurs due to simultaneously changing mass coefficients in the amplitude of such a decay.
We previously studied a similar decay mechanism in connection with the production of heavy quarkonia in \cite{apm2021,apm2023,apm2023a}.

The general structure of the tensor corresponding to the $W$-boson or quark loops in Fig.~\ref{fig11} in the case of two external virtual photons is the following:
\begin{equation}
\label{AB1}
T^{\mu\nu}_{Q,W}=A_{Q,W}(t)(g^{\mu\nu}(v_1v_2)-v_1^\nu v_2^\mu)+B_{Q,W}(t)[v_1^2v_2^\mu-v_1^\mu (v_1v_2)][v_2^2v_1^\nu-v_2^\nu(v_1v_2)],
\end{equation}
where  $t=\frac{M_H^2}{4m_Q^2}$ or $t=\frac{M_H^2}{4m_W^2}$, $m_W$ is the mass of $W$-boson,
$m_Q$ is the mass of heavy quark in the quark loop.
The structure functions $A_{Q,W}(t)$, $B_{Q,W}(t)$ can be obtained using an explicit expression for a loop integrals \cite{pp1,pp2,pp3}.
In the case where one photon is real, the only contribution comes from the structure function $A(t)$ which satisfies the dispersion relation:
\begin{equation}
\label{AB2}
A_W(t)=A_W(0)+\frac{t}{\pi}\int_1^\infty\frac{Im A(t') dt'}{t'(t'-t+i0)}.
\end{equation}

Imaginary part $Im A(t)$ can be calculated using the Mandelstam-Cutkosky rule \cite{t4}:
\begin{gather}
\label{AB3}
Im A_W=\frac{r_4^2}{64\pi}\frac{1}{t(4t-r_4^2)^2}
\Biggl[r_4^2\sqrt{t(t-1)}(r_4^2(2t+1)-4t-6)+ \\
4t(6-12t+r_4^2(2t+3)-r_4^4)arcsh(\sqrt{t-1})\Biggr], ~~~r_4=\frac{M}{M_W}.\nonumber 
\end{gather}
Since the mass parameter $r_4$ is very small, to calculate the function $A(t)$ itself from its imaginary part, one can use the expansion in $r_4$. In leading order by $r_4$ we get:
\begin{gather}
\label{AB4}
A_W(t)=\frac{r_4^2}{16\pi^2}\left[
2+\frac{3}{t}+\frac{3}{t^2}(2t-1)f^2(t)\right],\\
f(t)=
\begin{cases}
\arcsin\sqrt{t},~~~t\leq 1\\
\frac{i}{2}\left[\ln\frac{1-\sqrt{1-t^{-1}}}{1+\sqrt{1-t^{-1}}}-i\pi\right],~~~t>1
\end{cases}
.
\end{gather}

Since the contribution of the $W$-boson loop mechanism to the decay width significantly exceeds the others, it is convenient to separately present the corresponding formula for the decay width:
\begin{equation}
\label{AB5}
\Gamma_W=\frac{8\pi^3\alpha^4(r_2^2-1)ctg^2\theta_W M_Z^2}
{r_2^3M^4}\vert\tilde\psi(0)\vert^2 A_W^2
\bigl[3\vert g_{1W}\vert^2-\frac{1}{4}(r_2^2-1)^2 \vert g_{2W}\vert^2\bigr],   
\end{equation}
\begin{gather}
g_{1W}=(r_2^2-1)(1+\frac{7}{3}\omega_1+\frac{11}{3}\omega_2),~~
g_{2W}=2(1+\frac{7}{3}\omega_1+\frac{11}{3}\omega_2).
\end{gather}
We emphasize that this formula contains mass factors constructed from the observed particle masses. Thus, the coupling effects in the lepton system are taken into account despite their smallness.

Similarly to the $W$-loop mechanism, we can calculate the contribution of the quark loop mechanism shown in Fig.~\ref{fig11}. The imaginary part of the structure function $A_Q(t)$ has the form:
\begin{gather}
\label{AB6}
Im A_Q=\frac{r_5^2}{32\pi}\frac{1}{t(4t-r_5^2)^3}
\Biggl[3r_5^2\sqrt{t(t-1)}(4t-r_5^2)+ \\
4t(r_5^4+2r_5^2(1-2t)+8(t-1)t)arcsh(\sqrt{t-1})\Biggr], ~~~r_5=\frac{M}{m_Q}.\nonumber 
\end{gather}

After calculating the dispersion integral, the following result is obtained for the function $A_Q(t)$ in leading order in $r_5^2$ in the form:
\begin{gather}
\label{AB7}
A_Q(t)=\frac{r_5^2}{16\pi^2}\left[
\frac{1}{t}+\frac{(t-1)}{t^2}f^2(t)\right].
\end{gather}

Total decay width, taking into account quark loops, can then be obtained from formula \eqref{AB5}, making the following substitution:
\begin{equation}
\label{AB8}
g_{1W,2W}\to g_{1W,2W}\left(
1+\sum_{Q=c,b,t}\frac{24q_Q^2m_Q^2}{\cos^2{\theta_W}M_Z^2}\frac{A_Q}{A_W}
\right).
\end{equation}

Numerical values for the width of single dilepton state production are discussed below along with paired production processes.

\section{Pair production of leptonic bound states}

\begin{figure}[htbp]
\includegraphics[width=0.97\textwidth]{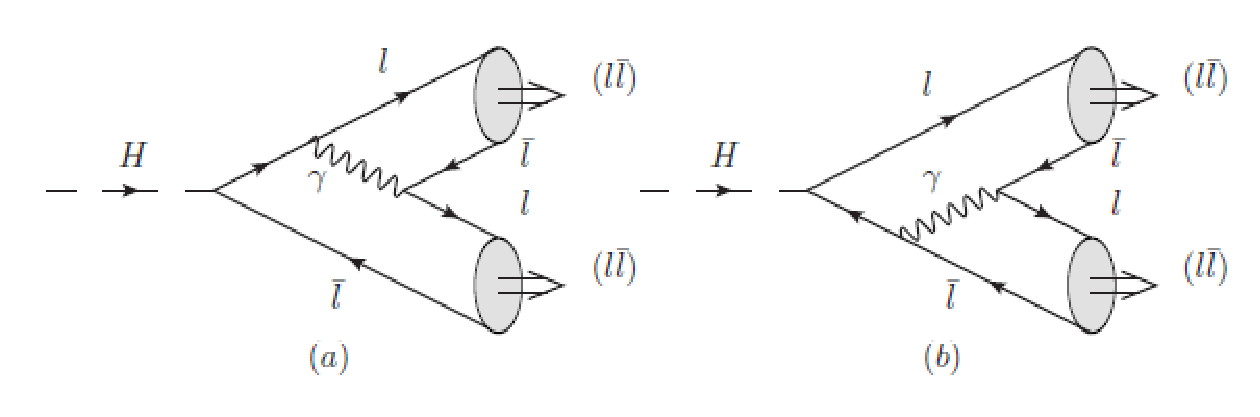}
\caption{Lepton - photon mechanism A of pair leptonium production in Higgs boson decay.}
\label{fig2}
\end{figure}

In addition to the processes of single production of bound states of leptons, we also consider processes of pair production of orto-positronium ($e^+ e^-$), orto-dimuonium ($\mu^+ \mu^-$), orto-ditauonium ($\tau^+ \tau^-$) in the Higgs boson decay. There are five different production mechanisms presented in Figs.~(\ref{fig2} - \ref{fig6}), which we consider below.

\subsection{Lepton - photon production mechanism A}
\label{subsection_mech_1}

The first lepton-photon production mechanism
can be separated into two stages. On the first stage two lepton - antilepton pair are produced. The first pair is produced in direct Higgs boson decay. After that lepton or antilepton can emit a photon or Z-boson, that produces another lepton - antilepton pair. On the second stage free leptons are combined in bound states.

The width of the decay process $H\to (l\bar l)_{S=1}+(l\bar l)_{S=1}$ is determined by an expression \eqref{dgamma}.
Four-momenta $p_1$, $p_2$ of one lepton-antilepton pair and $q_1$, $q_2$ of second pair can be expressed in terms of total $P$, $Q$ and relative $p$, $q$ 4-momenta in the form:
\begin{equation}
\label{kinematics}
p_{1,2} = \frac{1}{2} P \pm p, ~~~ q_{1,2} = \frac{1}{2} Q \pm q, ~~~(p P) = (q Q) = 0.
\end{equation}

Total amplitude of pair production ${\cal M}(P,Q)$ 
now contains two convolutions of the Higgs boson decay amplitude with the wave functions of two lepton bound states:
\begin{gather}
{\cal M}(P,Q)^{(A)} = 4\pi\alpha \Gamma_{Hl{\bar l}} \int \frac{d{\bf p}}{(2\pi)^3} \int \frac{d{\bf q}}{(2\pi)^3} 
Tr \Bigl\{ 
\Psi(p,P)\frac{-{\hat r}+{\hat p}_1 + m}{(-r+p_1)^2-m^2}\gamma^{\mu}\Psi(q,Q)\gamma^{\nu} \nonumber +  \\
\Psi(p,P)\gamma^{\mu}
\frac{{\hat r}-{\hat q}_1 + m}{(r-q_1)^2-m^2} \Psi(q,Q)\gamma^{\nu} \Bigr\} D^{\mu\nu}_\gamma (k_1), 
\label{1A1:eq1}
\end{gather}
where $\Gamma_{Hl{\bar l}}=m(\sqrt{2}G_F)^{1/2}$ is a vertex of the Higgs-lepton-antilepton interaction, $D^{\mu\nu}_\gamma (k_1)$ is a photon propagator, $k_1 = p_2+q_2$. 

Taking into account the law of transformation of the wave function of a bound state upon transition to a rest frame, we obtain wave functions in \eqref{1A1:eq1} in the form \cite{apm2021,apm2023}:
\begin{gather}
\Psi(p,P) = \frac{\psi_0({\bf p})}{\frac{\varepsilon(p)}{m}\frac{\varepsilon(p)+m}{2m}} 
\Bigl[ \frac{{\hat v}_1-1}{2} - {\hat v}_1\frac{p^2}{2m(\varepsilon(p)+m)} - \frac{{\hat p}}{2m} \Bigr] \times \nonumber \\
\frac{1}{2\sqrt{2}} {\hat \varepsilon}_1(P,S_z) (1+{\hat v}_1) 
\Bigl[ \frac{{\hat v}_1+1}{2} - {\hat v}_1\frac{p^2}{2m(\varepsilon(p)+m)} + \frac{{\hat p}}{2m} \Bigr], 
\label{1A1:eq2}
\end{gather}
\begin{gather}
\Psi(q,Q) = \frac{\psi_0({\bf q})}{\frac{\varepsilon(q)}{m}\frac{\varepsilon(q)+m}{2m}} 
\Bigl[ \frac{{\hat v}_2-1}{2} - {\hat v}_2\frac{q^2}{2m(\varepsilon(q)+m)} + \frac{{\hat q}}{2m} \Bigr] \times \nonumber \\
\frac{1}{2\sqrt{2}} {\hat \varepsilon}_2(Q,S_z) (1+{\hat v}_2) 
\Bigl[ \frac{{\hat v}_2+1}{2} - {\hat v}_2\frac{q^2}{2m(\varepsilon(q)+m)} - \frac{{\hat q}}{2m} \Bigr], 
\label{1A1:eq3}
\end{gather}
where we introduce two auxiliary 4-vectors $v_1=\frac{P}{M}$, $v_2=\frac{Q}{M}$, $\varepsilon(p)=\sqrt{{\bf p}^2+m^2}$, $m$ is the lepton mass. In equations \eqref{1A1:eq2}, \eqref{1A1:eq3} the projection operators on $(^3S_1)$ dilepton states are introduced. Projection operators are constructed in terms of the Dirac bispinors in the rest frame as follows:
\begin{equation}
\label{projection_operators}
\Pi_{S=1} = \frac{(1+\hat v_i)}{2\sqrt{2}} {\hat \varepsilon}_i(P,S_z).
\end{equation}

The denominators of propagators in amplitude \eqref{1A1:eq1} can be simplified due to the condition $M_H \gg m$ as follows:
\begin{gather}
\frac{1}{k_1^2}=\frac{1}{(p_2+q_2)^2} \approx \frac{4}{M_H^2},~~~
\frac{1}{(-r+p_1)^2-m^2} \approx \frac{1}{(r-q_1)^2-m^2} \approx \frac{2}{M_H^2}.
\label{1A1:eq4}
\end{gather}

The calculation of trace in \eqref{1A1:eq1} of the product of $\gamma$-factors and various simplifications are performed in the Form package \cite{form}.
Considering small factors $\frac{{\bf p}}{m}$, $\frac{{\bf q}}{m}$, we perform an expansion on them in the numerator of the amplitude, while preserving the second-order terms. Then the numerator of this amplitude takes the form:
\begin{gather}
N^{(A)} = \left(g_1^{(A)} g^{\alpha \beta}-g_2^{(A)} v_1^\beta v_2^\alpha\right) 
\varepsilon_1^\alpha(P,S_z) \varepsilon_2^\beta (Q,S_z),
\label{1A1:eq5}
\end{gather}
where the introduced functions are equal:
\begin{gather}
g_2^{(A)} = 1-\frac{1}{9} \omega_1^2,~~
g_1^{(A)} = - \frac{1}{2} + \frac{1}{2}r_2^2 - r_1 + \frac{2}{3}\omega_1 r_1 + \frac{1}{18} \omega_1^2 - \frac{1}{18} \omega_1^2 r_2^2 + \frac{1}{3} \omega_1^2 r_1.
\label{1A1:eq6}
\end{gather}
Special relativistic parameters $\omega_1$, $\tilde\psi(0)$ take into account relativistic corrections connected with the relative motion of leptons. The calculation of such parameters is discussed in details below in section \ref{sec_relativistic_corrections}. After all transformations, the production amplitude \eqref{1A1:eq1} takes the form:
\begin{gather}
{\cal M}(P,Q)^{(A)} = 
\frac{64 \pi \alpha m M (\sqrt{2}G_F)^{1/2} {\tilde \psi}^2(0)}{M_H^4}(g_1^{(A)} g^{\alpha \beta}-
g_2^{(A)} v_1^\beta v_2^\alpha \varepsilon_1^\alpha(P,S_z) \varepsilon_2^\beta (Q,S_z). ~~~
\label{prod_amp_1}
\end{gather}

\subsection{Lepton - Z-boson production mechanism B} 
\label{subsection_mech_2}
\begin{figure}[htbp]
\includegraphics[width=0.97\textwidth]{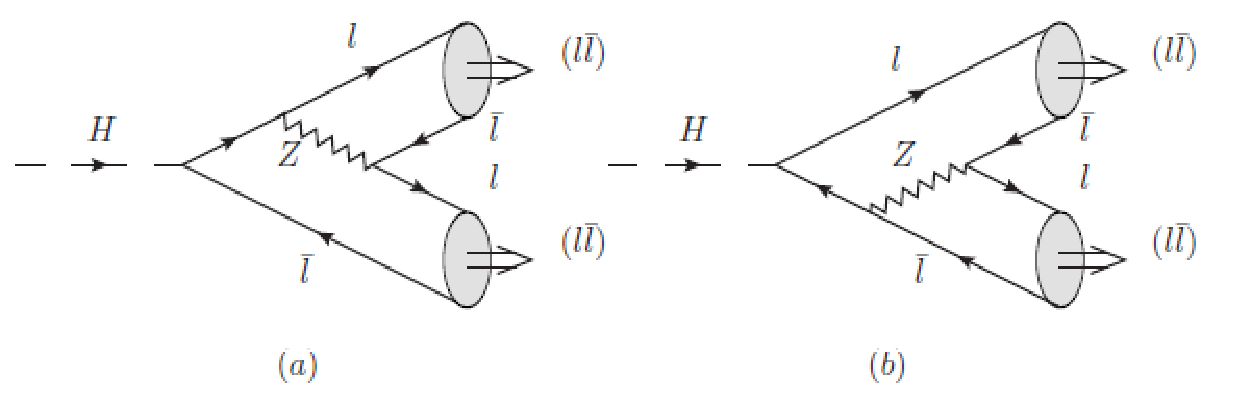}
\caption{Lepton - $Z$-boson mechanism B of pair leptonium production in the Higgs boson decay.}
\label{fig3}
\end{figure}

The second mechanism of pair production differs from the previous one by replacing the photon with a $Z$-boson.
The decay amplitudes of the Higgs boson are shown in Fig.~\ref{fig3}.
The total amplitude of the process is determined in this case by the following expression:
\begin{gather}
{\cal M}(P,Q)^{(B)} = -i \Gamma_{Hl{\bar l}} \int \frac{d{\bf p}}{(2\pi)^3} \int \frac{d{\bf q}}{(2\pi)^3}
Tr\Bigl\{ 
\Psi(p,P)\frac{-{\hat r}+{\hat p}_1 + m}{(-r+p_1)^2-m^2}\Gamma^{\mu}_{Z l{\bar l}} \Psi(q,Q)\Gamma^{\nu}_{\gamma l{\bar l}} \nonumber +  \\
\Psi(p,P)\Gamma^{\mu}_{\gamma l{\bar l}}
\frac{{\hat r}-{\hat q}_1 + m}{(r-q_1)^2-m^2} \Psi(q,Q)\Gamma^{\nu}_{{Z l{\bar l}}} 
\Bigr\} D^{\mu\nu}_Z (k_2), 
\label{1A2:eq1}
\end{gather}
where $D^{\mu\nu}_Z (k_2)$ is the $Z$-boson propagator, $k_2 = p_2+q_2$. The $Z$-boson - lepton-antilepton interaction vertex is determined by:
\begin{gather}
\Gamma^{\mu}_{Z l{\bar l}} = \frac{e}{\sin{2\theta_W}} \gamma_\mu \left[ \frac{1}{2}(1-\gamma^5)+2\sin^2{\theta_W} \right].
\label{zllvertex}
\end{gather}
When transforming the denominators of the $Z$-boson and lepton propagators, we also neglect the relative momenta $p$ and $q$ and write them in the form:
\begin{gather}
\frac{1}{k_2^2-M_Z^2}=\frac{1}{(p_2+q_2)^2} \approx \frac{4}{M_H^2 - 4 M_Z^2},~
\frac{1}{(-r+p_1)^2-m^2} \approx \frac{1}{(r-q_1)^2-m^2} \approx \frac{2}{M_H^2}.  
\label{1A2:eq2}
\end{gather}
After the trace calculation the numerator of total amplitude has the same general structure as \eqref{1A1:eq5}:
\begin{gather}
N^{(B)}=\left( g_1^{(B)} g^{\alpha \beta}-
g_2^{(B)} v_1^\beta v_2^\alpha \right) 
\varepsilon_1^\alpha(P,S_z) \varepsilon_2^\beta (Q,S_z),
\label{1A2:eq3}
\end{gather}
and the functions $g_1^{(B)}$, $g_2^{(B)}$ in the considered approximation are equal:
\begin{gather}
g_2^{(B)} =2(1-2a_z+2 a_z^2)(9-\omega_1^2), \\ 
g_1^{(B)} =(1-2a_z+2a_z^2)
(3+\omega_1)(-3+\omega_1 -6 r_1+6\omega_1 r_1 +3r_2^2-\omega_1 r_2^2), ~~~~~~~
\label{1A2:eq4}
\end{gather}
where $a_z=2\sin^2\theta_W$. 
The resulting expression for total amplitude of this decay mechanism has the form:
\begin{gather}
{\cal M}(P,Q)^{(B)} =  
\frac{64 \pi \alpha m M (\sqrt{2}G_F)^{1/2} {\tilde \psi}^2(0)}{M_H^2 (M_H^2-4M_Z^2)  \sin^2 2\theta_W}
(g_1^{(B)} g^{\alpha \beta}-
g_2^{(B)} v_1^\beta v_2^\alpha)\varepsilon_1^\alpha(P,S_z) \varepsilon_2^\beta (Q,S_z).~~~ 
\label{prod_amp_2}
\end{gather}

\subsection{Lepton - photon production mechanism C} 
\label{subsection_mech_3}
\begin{figure}[htbp]
\includegraphics[width=0.97\textwidth]{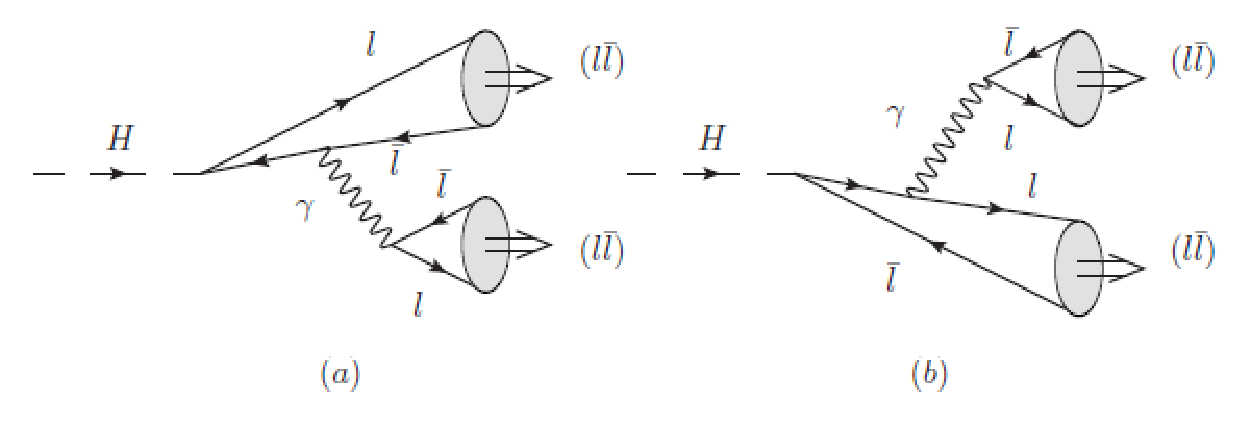}
\caption{Lepton - photon mechanism C of pair leptonium production in the Higgs boson decay.}
\label{fig4}
\end{figure}

There are two more amplitudes in the lepton-photon decay mechanism, which are shown in Fig.~\ref{fig4}.
Their structure is different from the previous ones in subsection \ref{subsection_mech_1}, so we consider them separately.
The first difference is that here there is a product of two different traces. The second difference is that the denominator of the photon propagator does not contain the large mass of the Higgs boson or $Z$-boson.
The denominator of the photon propagator $k_3^2$ ($k_3=q_1+q_2$), which is equal to the mass of the bound state of leptons, can be considered as a factor that increases the contribution of such a mechanism to the overall decay amplitude.
Let us present here total amplitude of the process, which is obtained according to the same rules as the previous one in Fig.~\ref{fig2}:
\begin{gather}
\label{1A3:eq1}
{\cal M}(P,Q)^{(C)} = -i \Gamma_{Hl{\bar l}} \int \frac{d{\bf p}}{(2\pi)^3} \int \frac{d{\bf q}}{(2\pi)^3}
Tr\Bigl\{ 
\Psi(p,P)\frac{(-{\hat r}+{\hat p}_1 + m)}{(-r+p_1)^2-m^2}\Gamma^{\mu}_{\gamma l{\bar l}}  +  \\
\Psi(p,P)\gamma^{\mu}
\frac{({\hat r}-{\hat q}_1 + m)}{(r-p_2)^2-m^2} \Bigr\} 
Tr \Bigl\{  \Psi(q,Q)\gamma^{\nu}\Bigr\} D^{\mu\nu}_\gamma (k_3). \nonumber 
\end{gather}
Now the mass of the Higgs boson is contained in only one of the denominators in the amplitude \eqref{1A3:eq1} in the form:
\begin{gather}
\label{1A3:eq2}
\frac{1}{k_3^2}=\frac{1}{q_1+q_2} \approx \frac{1}{M^2},~~~
\frac{1}{(-r+p_1)^2-m^2} \approx \frac{1}{(r-p_2)^2-m^2} \approx \frac{2}{M_H^2}.
\end{gather}
The numerator of this amplitude is also a convolution of the ortho-dileptonic polarization vectors with a tensor, which is expressed through two new functions in the form:
\begin{gather}
\label{1A3:eq3}
N^{(C)} = \left( g_1^{(C)} g^{\alpha \beta}-
g_2^{(C)} v_1^\beta v_2^\alpha\right) 
\varepsilon^\alpha(P,S_z) \varepsilon^\beta (Q,S_z),
\end{gather}
\begin{gather}
\label{1A3:eq4}
g_1^{(C)}=r_2^2-1 -2 r_1 +\frac{4}{3}\omega_1 r_1+\frac{1}{9} \omega_1^2-\frac{1}{9}\omega_1^2 r_2^2+\frac{2}{3}\omega_1^2r_1,~~~
g_2^{(C)}=2-\frac{2}{9}\omega_1^2.
\end{gather}
Total amplitude of the production of a pair of dileptonia in this mechanism is represented as follows:
\begin{gather}
\label{prod_amp_3}
{\cal M}(P,Q)^{(C)} = 
-\frac{8 (4\pi\alpha)^{3/2} m  {\tilde \psi}^2(0)}{M_H^2 M_Z\sin 2\theta_W} 
(g_1^{(C)} g^{\alpha \beta}-
g_2^{(C)} v_1^\beta v_2^\alpha \varepsilon_1^\alpha(P,S_z) \varepsilon_2^\beta (Q,S_z). 
\end{gather}

\subsection{Lepton - Z-boson production mechanism D} 
\label{subsection_mech_4}
\begin{figure}[htbp]
\includegraphics[width=0.97\textwidth]{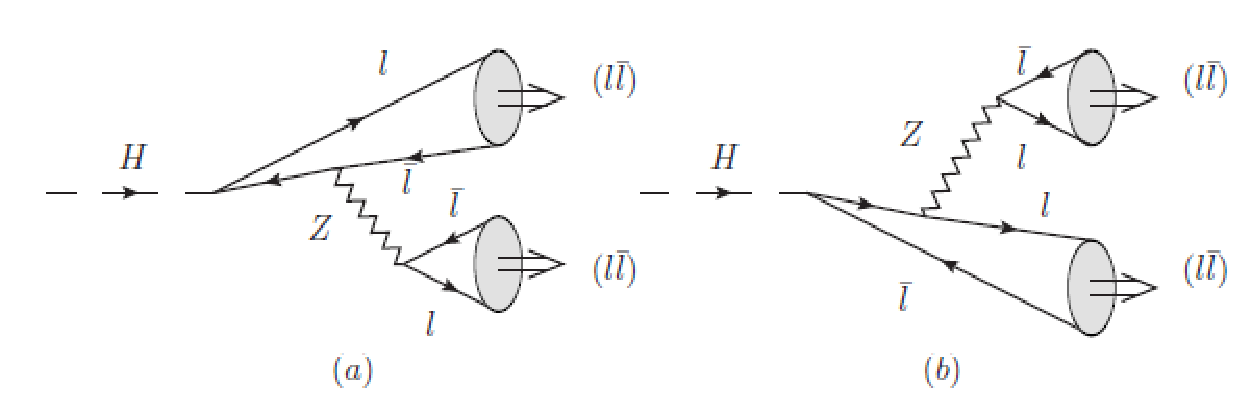}
\caption{Lepton - $Z$-boson mechanism D of pair leptonium production in the Higgs boson decay.}
\label{fig5}
\end{figure}

Replacing the photon with $Z$-boson, we obtain two more pair production amplitudes shown in Fig.~\ref{fig5}
But unlike the previous photon production of a bound state of leptons, now the denominator of the $Z$-boson propagator still contains a large mass of the $Z$-boson, which may mean the suppression of this mechanism compared to the lepton-photon one.
However, we have also included this amplitude for completeness.

The amplitude for the production of a leptonium pair takes the form:
\begin{gather}
{\cal M}(P,Q)^{(D)} = -i \Gamma_{Hl{\bar l}} \int \frac{d{\bf p}}{(2\pi)^3} \int \frac{d{\bf q}}{(2\pi)^3}
Tr\Bigl\{ 
\Psi(p,P)\frac{-{\hat r}+{\hat p}_1 + m}{(-r+p_1)^2-m^2}\Gamma^{\mu}_{Z l{\bar l}}  +   \nonumber  \\
\Psi(p,P)\Gamma^{\mu}_{Z l{\bar l}}
\frac{{\hat r}-{\hat q}_1 + m}{(r-p_2)^2-m^2} \Bigr\} 
Tr \Bigl\{  \Psi(q,Q)\Gamma^{\nu}_{{Z l{\bar l}}} \Bigr\} D^{\mu\nu}_Z (k_4), 
\label{1A4:eq1}
\end{gather}
where $k_4=q_1+q_2$.

The propagator denominators contain two large masses of the $Z$-boson and the Higgs boson:
\begin{gather}
\frac{1}{k_4^2 - M_Z^2}=\frac{1}{(q_1+q_2)^2-M_Z^2} \approx \frac{1}{M^2-M_Z^2},~~ \nonumber \\
\frac{1}{(-r+p_1)^2-m^2} \approx \frac{1}{(r-p_2)^2-m^2} \approx \frac{2}{M_H^2}. 
\label{1A4:eq2}
\end{gather}
After calculating two traces in the numerator \eqref{1A4:eq1} we obtain:
\begin{gather}
N^{(D)} = \left( g_1^{(D)} g^{\alpha \beta}-
g_2^{(D)} v_1^\beta v_2^\alpha \right) 
\varepsilon_1^\alpha(P,S_z) \varepsilon_2^\beta (Q,S_z),
\label{1A4:eq3}
\end{gather}
\begin{gather}
g_2^{(D)} =2(-1 + 2 a_z) (9 - \omega_1^2),  \\
g_1^{(D)} =  (-1 + 2 a_z)
(3 + \omega_1) (-3 + \omega_1 - 6 r_1 + 6 \omega_1 r_1 + 3 r_2^2 - \omega_1 r_2^2).\nonumber
\label{1A4:eq4}
\end{gather}
Then the production amplitude \eqref{1A4:eq1} has the same general structure as previous amplitudes:
\begin{gather}
{\cal M}(P,Q)^{(D)} = 
\frac{16 \pi \alpha m M (\sqrt{2}G_F)^{1/2} {\tilde \psi}^2(0)}{M_H^2 (M^2-M_Z^2) \sin^2(2\theta_W)}
( g_1^{(D)} g^{\alpha \beta}-
g_2^{(D)} v_1^\beta v_2^\alpha)\varepsilon_1^\alpha(P,S_z) \varepsilon_2^\beta (Q,S_z). 
\label{prod_amp_4}
\end{gather}

\subsection{ZZ-boson production mechanism E}
\label{subsection_mech_5}
\begin{figure}[htbp]
\includegraphics[width=0.97\textwidth]{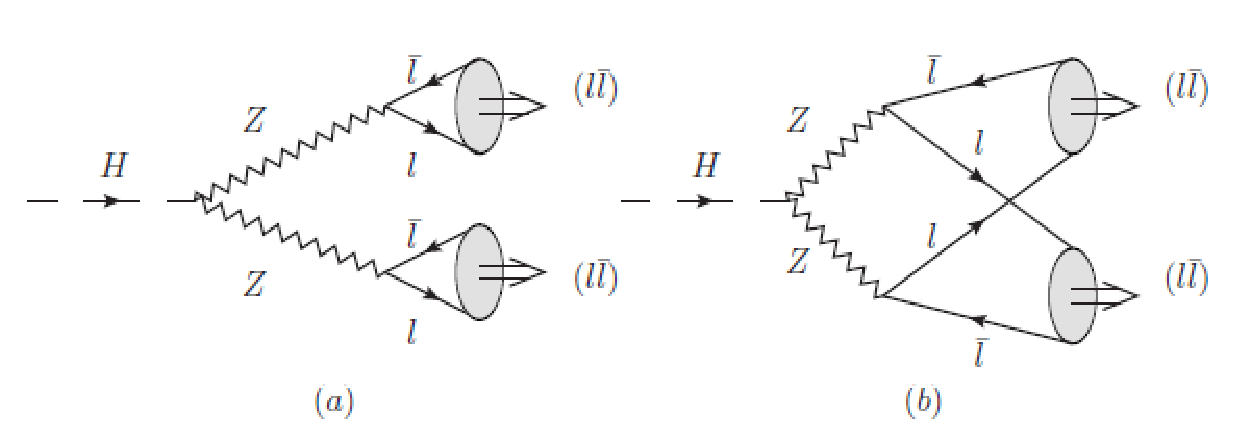}
\caption{Pair leptonium ZZ-boson production mechanism in the Higgs boson decay.}
\label{fig6}
\end{figure}

The production of a pair of bound states of particles (quarks, leptons) can occur due to $ZZ$-mechanism shown in Fig.~\ref{fig6}. A pair of $Z$-bosons arises directly from the decay of the Higgs boson. 
The required vertex functions have the form:
\begin{equation}
\label{ZZ}
\Gamma^{\alpha\beta}=\frac{2e}{\sin 2\theta_W} M_Z g^{\alpha\beta},~~~\Gamma^\mu_{Zl\bar l}=\frac{e}{\sin 2\theta_W }
\gamma^{\mu}\left [\frac{1}{2}(1-\gamma_5)+2\sin^2 \theta_W
\right].
\end{equation}

In this case, there are direct (a) and crossed (b) amplitudes (see Fig.\ref{fig6}):
\begin{gather}
{\cal M}(P,Q)^{(E1)}=\Gamma^{\mu'\nu'} \int \frac{d{\bf p}}{(2\pi)^3} \int \frac{d{\bf q}}{(2\pi)^3}
Tr \Bigl\{ \Psi(p,P)\Gamma^{\mu}_{Z l{\bar l}} \Bigr\} 
Tr \Bigl\{ \Psi(q,Q)\Gamma^{\nu}_{Z l{\bar l}} \Bigr\} D^{\mu\mu'}_Z (k_5) D^{\nu\nu'}_Z (k_6), 
\label{1A5:eq1}
\end{gather}
where $k_5=p_1+p_2$, $k_6=q_1+q_2$.
\begin{gather}
{\cal M}(P,Q)^{(E2)}=\Gamma^{\mu'\nu'} \int 
\frac{d{\bf p}}{(2\pi)^3} \int \frac{d{\bf q}}{(2\pi)^3}
Tr \Bigl\{ \Psi(p,P)\Gamma^{\mu}_{Z l{\bar l}} \Psi(q,Q)\Gamma^{\nu}_{Z l{\bar l}} \Bigr\} 
D^{\mu\mu'}_Z (k_7) D^{\nu\nu'}_Z (k_8), 
\label{1A5:eq2}
\end{gather}
where $k_7=p_1+q_1$, $k_8=p_2+q_2$.

By choosing the $Z$-boson propagator in the unitary gauge, we can represent the total amplitude as follows:
\begin{gather}
{\cal M}(P,Q)^{(E)}=\frac{8(4\pi\alpha)^{3/2}M}{M_z^3\sin^3\theta_W}
\tilde\psi^2(0)\Biggl\{(\varepsilon_1\varepsilon_2)\left[g_1^{E,dir}+
\frac{r_3^4}{(r_3^2-\frac{r^2_2}{4})^2}g_1^{E,cr}\right]-\\
(\varepsilon_1 v_2)(\varepsilon_2 v_1)\left[g_2^{E,dir}+
\frac{r_3^4}{(r_3^2-\frac{r^2_2}{4})^2}g_2^{E,cr}\right]
\Biggr\}, \nonumber
\label{zztotal}
\end{gather}
\begin{gather}
g_1^{E,dir}=\frac{1}{4}+\frac{1}{6}+a_z+\frac{2}{3}a_z\omega_1+a_z^2+
\frac{2}{3}a_z^2\omega_1,~~~g_2^{E,dir}=0,   \\
g_1^{E,cr}=-\frac{1}{4}-\frac{r_2^4}{128r_3^4}-\frac{r_2^4\omega_1}{192r_3^4}-\frac{r_2^4 a_z\omega_1}{48r_3^4}-\frac{r_2^4 a^2_z\omega_1}{48r_3^4}+\frac{r_2^2}{16r_3^2}+\frac{r_2^2\omega_1}{24r_3^2}+\frac{r_2^2a_z\omega_1}{6r_3^2}+\frac{r_2^2a^2_z\omega_1}{6r_3^2}- \nonumber  \\
\frac{1}{6}\omega_1-\frac{1}{2}a_z-\frac{1}{3}\omega_1 a_z-
\frac{1}{2}a_z^2-\frac{1}{3}\omega_1a_z^2, \\
g_2^{E,cr}=-\frac{r_2^2}{64r_3^4}-\frac{r_2^2\omega_1}{96r_3^4}-
\frac{r_2^2\omega_1 a_z}{24r_3^4}-\frac{r_2^2\omega_1 a^2_z}{24r_3^4}+\frac{1}{8r_3^2}+\frac{\omega_1}{12r_3^2}+
\frac{\omega_1 a_z}{3r_3^2}+\frac{\omega_1 a_z^2}{3r_3^2}. 
\label{zzt1}
\end{gather}

The indices dir and cr denote the contributions of the direct and crossed interaction amplitudes in Fig.\ref{fig7}.

\subsection{Quark loop and W-boson loop production mechanism}
\label{subsection_mech_pair_loop}

\begin{figure}[htbp]
\includegraphics[width=0.97\textwidth]{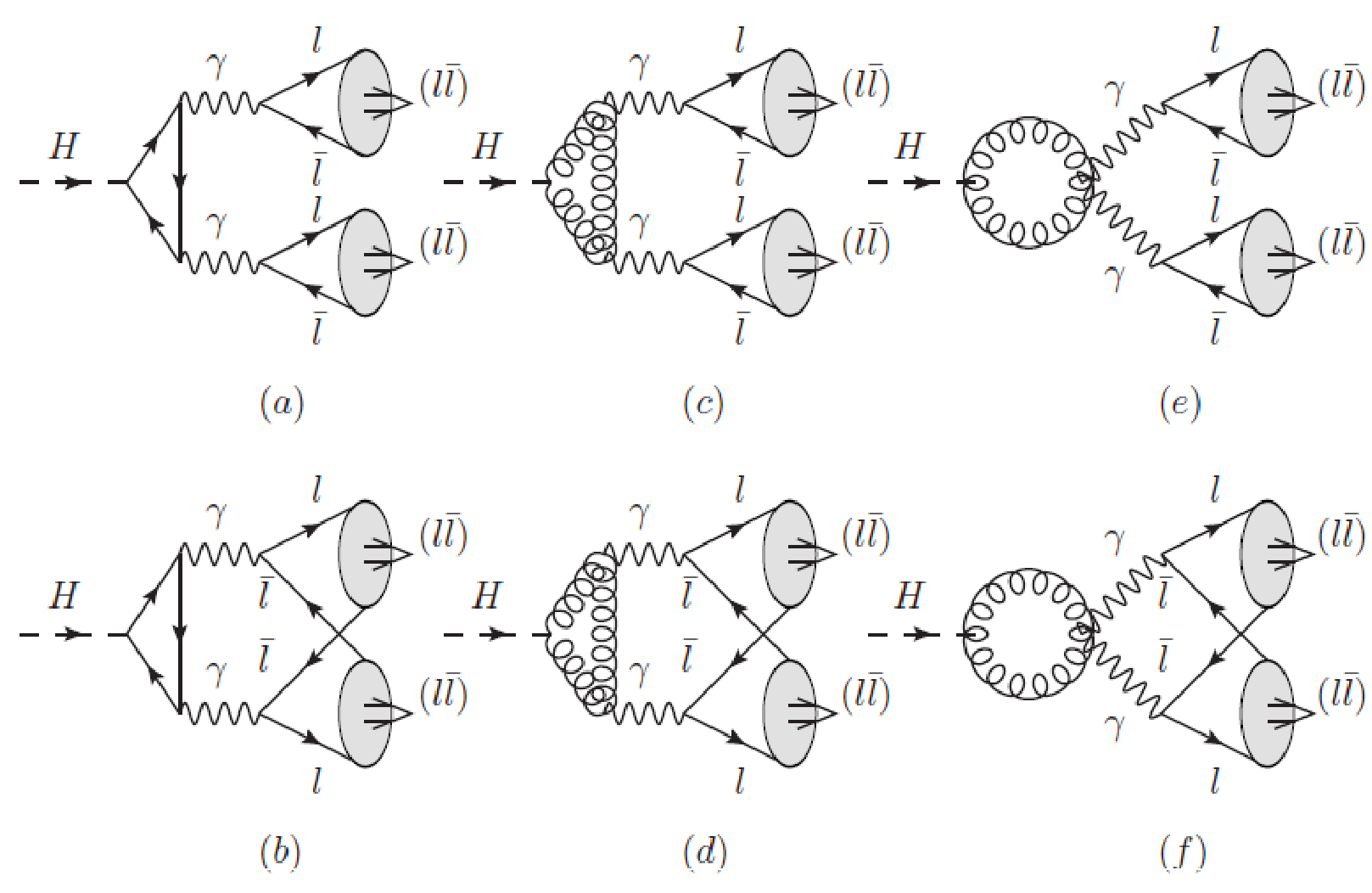}
\caption{Pair leptonium production mechanism in the Higgs boson decay with quark and $W$-boson loop.}
\label{fig7}
\end{figure}

From studies of paired meson production in the decay of the Higgs boson, it is known that the $W$-boson and quark loop mechanisms can make a significant contribution to the overall decay width.
The corresponding amplitudes are presented in Fig.~\ref{fig7}.
In each column here we have direct and crossed amplitudes. In this case, the contribution of the crossed amplitude is suppressed by the degree of small mass parameter, which arises from the denominators of virtual photons.

Since in the case of boson and quark loops the amplitude includes tensor \eqref{AB1}, we can immediately obtain total contribution of all loops in the decay amplitude:
\begin{gather}
{\cal M}(P,Q)^{(loop)}=
\frac{64\sqrt{4\pi\alpha}\pi^2\alpha^2 M_Z ctg\theta_W}{M^3r_2^4}\tilde\psi^2(0)\Biggl\{
(\varepsilon_1\varepsilon_2)\Bigl(g_1^{W}+\sum_{Q=c,b,t}\frac{24 m_Q^2}{M_Z^2\cos^2\theta_W}g_1^{Q}\Bigr)- \nonumber \\
(\varepsilon_1 v_2)(\varepsilon_2 v_1)
\Bigl(g_2^W+\sum_{Q=c,b,t}\frac{24 m_Q^2}{M_Z^2\cos^2\theta_W}g_2^{Q}\Bigr)
\Biggr\},
\label{WQtotal}
\end{gather}
\begin{gather}
g_1^W=A^W(\frac{1}{3}\omega_1-\frac{1}{8}r_2^2-\frac{1}{12}\omega_1 r_2^2-r_2^4-\frac{2}{3}\omega_1r_2^4+\frac{1}{2}r_2^6+\frac{1}{3}\omega_1 r_2^6)+ \\
B^W(-\frac{1}{3}\omega_1 r_2^2+\frac{1}{8}r_2^4+\frac{1}{6}\omega_1 r_2^4-\frac{1}{32}r_2^6-\frac{1}{48}\omega_1 r_2^6),  \\
g_2^W=A^W(\frac{1}{4}+\frac{1}{6}\omega_1+r_2^4+\frac{2}{3}\omega_1 r_2^4)+B^W(-\frac{2}{3}\omega_1+\frac{1}{3}\omega_1 r_2^2-\frac{17}{16}r_2^4-\frac{17}{24}\omega_1r_2^4).
\label{WQtotal1}
\end{gather}
The coefficients $g_1^Q$, $g_2^Q$ are obtained from $g_1^W$, $g_2^W$ by replacing $A^W\to A^Q$, $B^W\to B^Q$.

All decay amplitudes presented above, obtained within the framework of various mechanisms, have the same structure and are expressed in terms of the functions $g_1$ and $g_2$. Accordingly, total decay amplitude has the same structure and contains total functions $g_1^{tot}$ and $g_2^{tot}$ with corresponding coefficients.

Total width of the decay of the Higgs boson into a pair of bound states of leptons can be obtained from \eqref{AB6} in terms of the functions $g_1^{tot}$, $g_2^{tot}$ in the form:
\begin{gather}
\Gamma=\frac{2048\pi^4\alpha^5 r_3^2\sqrt{(\frac{r_2^2}{4}-1)}ctg^2\theta_W}{ M^5 r_2^{10}}\vert\tilde\psi(0)\vert^4
\Bigl[|g_1^{tot}|^2(3-r_2^2+\frac{1}{4}r_2^4)+  \nonumber  \\
|g_2^{tot}|^2(r_2^4-\frac{1}{2}r_2^6+\frac{1}{16}r_2^8)+(g_2^{tot} g_1^{\ast~tot}+g_1^{tot} g_2^{\ast~tot})(-r_2^2+\frac{3}{4}r_2^4-\frac{1}{8}r_2^6)\Bigr],
\label{WQtotal2}
\end{gather}
where the general factor corresponds to amplitude \eqref{WQtotal} so that the functions $g^W_1$ and $g^W_2$ are included in the complete functions $g_1^{tot}$ and $g_2^{tot}$ with unit coefficients.

\section{Relativistic corrections} 
\label{sec_relativistic_corrections}

Relativistic corrections represent an important class of corrections for the production of bound quark states. In the case of the production of lepton bound states, such corrections are small, since they are determined by the value of the fine structure constant. Taking them into account is connected with increasing the accuracy of calculating the decay width.

When calculating decay widths, we take into account relativistic corrections of several types. The first group of relativistic corrections is connected with the law of transformation of the wave function of a bound state into a rest frame and the momenta of the relative motion of leptons in the decay amplitude. The second group is determined by the Breit potential in the wave function of the bound state in the rest frame.

\subsection{Relativistic parameters in production amplitude} 
\label{sec_relativistic_parameters}

To calculate relativistic corrections to the $H$-boson decay amplitude, we use the same approach as in our previous works \cite{apm2021,apm2023}. After calculating the trace in the numerator of the amplitudes, corrections of a certain order of smallness in the relative momentum of leptons in the bound state are identified. We take into account second-order corrections in ${\bf p}^2/m^2$, ${\bf q}^2/m^2$ and introduce an expansion of the form:
\begin{gather}
\frac{\vert{\bf p}\vert}{2m} =\sum_{n=1}^{\infty} \left( \frac{\varepsilon(p)-m}{\varepsilon(p)+m} \right)^{n+\frac{1}{2}},~~~
\frac{\vert{\bf q}\vert}{2m} =\sum_{n=1}^{\infty} \left( \frac{\varepsilon(q)-m}{\varepsilon(q)+m} \right)^{n+\frac{1}{2}}.
\label{razl}
\end{gather}
This expansion emphasizes that the relative momenta of leptons are small compared to their mass. In quantum electrodynamics, the form of the wave function of a bound state in the initial approximation is known analytically. It follows from this that $\vert{\bf p}\vert\sim W=\mu\alpha$.

Using expansion \eqref{razl}, relativistic corrections in $p$ can be represented as the following characteristic integrals in momentum space:
\begin{gather}
I^{(i)} = \int \frac{d{\bf p}}{(2\pi)^3} \frac{\varepsilon(p)+m}{2\varepsilon(p)} \psi^C_{1S}({\bf p}) \left( \frac{\varepsilon(p)-m}{\varepsilon(p)+m} \right)^{i},
\label{2:eq1}
\end{gather}
\begin{equation}
\label{wave_function_1S_p}
\psi^C_{1S}(p) = \frac{8\sqrt{\pi} W^{5/2}}{(p^2+W^2)^2},~~~W=\frac{m}{2} \alpha.
\end{equation}

In the amplitudes presented earlier, relativistic effects are expressed in terms of special parameters, which are determined by the integral ratios \eqref{2:eq1}:
\begin{equation}
\label{omega}
\omega_1=\frac{I^{(1)}}{I^{(0)}},~~~\omega_2=\frac{I^{(2)}}{I^{(0)}},~~~I^{(0)}=\tilde\psi(0)=\int \frac{d{\bf p}}{(2\pi)^3} \frac{\varepsilon(p)+m}{2\varepsilon(p)} \psi^C_{1S}({\bf p}).
\end{equation}

After this, when calculating values \eqref{omega}, we perform an expansion in $p/m$ taking into account terms of the second degree.
All nonrelativistic expansions in $\vert{\bf p}\vert/m$ that we perform assume that $\vert{\bf p}\vert$ is small compared to $m$. Therefore, although integrals \eqref{omega} are formally convergent, when calculating them it is necessary to take the value of the momentum at the pole of the wave function. Proceeding in this way, we obtain the following results when calculating relativistic corrections:
\begin{gather}
\tilde\psi(0)=\int \frac{d{\bf p}}{(2\pi)^3}\psi^C_{1S}({\bf p})\left[1-\frac{{\bf p}^2}{4m^2}\right]=\psi^C_{1S}(0)\left[1+\frac{3}{4}\alpha^2\right], \\
I^{(1)}=\int \frac{d{\bf p}}{(2\pi)^3}\psi^C_{1S}({\bf p})\frac{{\bf p}^2}{4m^2}=-\frac{3}{4}\alpha^2\psi^C_{1S}(0),~~~
\omega_1=-\frac{3}{4}\alpha^2.
\label{omega1}
\end{gather}
The parameters $\omega_i$ for $i>1$ are proportional to higher powers of $\alpha$ than \eqref{omega1} and are omitted from now on.
The results \eqref{omega1} are obtained using the following expression \cite{ibk1994,apm1998}:
\begin{gather}
\int \frac{d {\bf p}}{(2\pi)^3} \frac{{\bf p}^2}{4m^2} \psi^C_n({\bf p}) = 
-\frac{[3+2(n-1)(n+1)]}{4n^2} \alpha^2 \psi^C_n(0).
\label{I0_v2_3}
\end{gather}

\subsection{Relativistic corrections to the wave function} \label{sec_relativistic_corrections_wave_function}

Wave function \eqref{wave_function_1S_p} was obtained from Schr\"odinger equation solution with Coulomb potential. In our previous works we had estimated relativistic correction to wave function by accounting corresponding terms in quark-antiquark potential \cite{apm2021,apm2023,apm2023a}. In present research we also can estimate leading order relativistic correction to wave function as follows \cite{sgk1,lepage1,lepage2,khriplovich,penin}:
\begin{equation}
\label{psicor}
\psi^{(1)}_{1S}(0) = \int {\tilde G}_{1S} (0,{\bf r}) \Delta V ({\bf r}) \psi^{(0)}_{1S}({\bf r}) d {\bf r},
\end{equation}
where ${\tilde G}_{1S}$ is the reduced Coulomb Green's function, $\psi^{(0)}_{1S}$ is defined by \eqref{wave_function_1S_p}, $\Delta V$ is the perturbative potential.

When accurately calculating the corresponding corrections to the wave function, it is necessary to use the Green's function with two non-zero arguments:
\begin{gather}
{\tilde G}_{1S}({\bf r}_1,{\bf r}_2) = -\frac{\mu W}{\pi} e^{-(x_> + x_<)} \Bigl(\frac{1}{2 x_>} - \ln(2 x_>) - \ln(2 x_<) + Ei(2 x_<) + \\
\frac{7}{2} - 2 \gamma_E - (x_> +x_<) + \frac{1 - e^{2 x_<}}{2 x_<} \Bigr) , \nonumber
\label{g1s}
\end{gather}
where $x_> = max(x_1,x_2)$, $x_< = min(x_1,x_2)$, $x_{1.2} = W r_{1,2}$, $\gamma_E = 0.577216\dots$ is Euler constant. When calculating corrections to the wave function at zero, they are determined by the Green's function with one zero argument:
\begin{gather}
{\tilde G}_{1S}({\bf r},0) = \frac{\mu W}{4 \pi} \frac{e^{-x}}{x} (4 x (\ln(2 x) + \gamma_E) + 4 x^2 - 10 x - 2),
\label{g1s0}
\end{gather}
where $x=W r$, $\gamma_E$ is the Euler constant.

The Breit potential contains several terms, the contribution of which must be calculated according to formula \eqref{psicor}:
\begin{gather}
\Delta V_1 = \frac{\pi \alpha}{m^2} \delta({\bf r})-\frac{{\bf p}^4}{4m^3},
\label{dv1}
\end{gather}
\begin{gather}
\Delta V_2 = -\frac{\alpha}{2 m^2}\frac{1}{r} 
\left( {\bf p}^2 + \frac{{\bf r}({\bf r}{\bf p}){\bf p}}{r^2} \right),
\label{dv2}
\end{gather}
\begin{gather}
\Delta V_3 = 
\frac{\pi\alpha}{m^2}\left[ \frac{7}{3}{\bf S}^2-2 \right]\delta({\bf r}).
\label{dv3}
\end{gather}

It is convenient to start calculating corrections to the wave function at zero with the $\delta$ term of the potential $\Delta V_1$:
\begin{gather}
\psi^{(1)}_{1S}(0) = {\tilde G}_{1S} (0,0) \frac{\pi \alpha}{m^2} \psi^{C}_{1S}(0),
\label{delta}
\end{gather}
where $\tilde G_{1S}(0,0)$ represents a divergent expression.
Along with the coordinate representation, it is useful to use the momentum representation when calculating corrections. In momentum representation this correction is determined by the divergent integral in the form:
\begin{gather}
\psi^{(1)}_{1}(0)=\frac{\pi \alpha}{m^2}\psi^{C}_{1S}(0)
\int\tilde G_{1S}({\bf q},{\bf p})\frac{d{\bf q}}{(2\pi)^3}
\frac{d{\bf p}}{(2\pi)^3},
\label{delta1}
\end{gather}
\begin{gather}
\tilde G_{1S}({\bf p},{\bf q})=-\frac{64\pi}{\alpha W^4}
\left[
\frac{\pi^2W^5\delta({\bf p}-{\bf q})}{4({\bf p}^2+W^2)}+
\frac{W^6}{4({\bf p}^2+W^2)({\bf p}-{\bf q})^2({\bf q}^2+W^2)}+
R({\bf p},{\bf q})\right].
\label{delta2}
\end{gather}

In parallel with this correction, let us also consider the contribution from the second term of operator \eqref{dv1}:
\begin{gather}
\psi^{(1)}_{2}(0)=-\frac{1}{4m^3}\int \tilde G_{1S}(0,{\bf r})\hat{\bf p}^4\psi^C_{1S}(r)d{\bf r})= \nonumber~\\
-\frac{\pi\alpha}{m^2}\psi^C_{1S}(0)\int \tilde G({\bf q},{\bf p})\frac{d{\bf p}d{\bf q}}{(2\pi)^6}\left[1-\frac{W^2(2p^2+W^2)}{(p^2+W^2)^2}\right].
\label{delta3}
\end{gather}
From expressions \eqref{delta1} , \eqref{delta3} it follows that the divergent terms in the matrix elements cancel each other so that the contribution of the operator from \eqref{dv1}  will be finite. After calculating three integrals with functions from \eqref{delta2} , we obtain the following summary result:
\begin{gather}
\Delta\psi^{(1)}(0)=\psi^C_{1S}(0)[-\frac{63}{128}\alpha^2].
\label{delta4}
\end{gather}

Let us now consider one more divergent term in $\psi(0)$, which is determined by the first part of operator \eqref{dv2}:
\begin{gather}
\Delta V^{(1)}_2=-\frac{\alpha}{2m^2r}\hat{\bf p}^2.
\label{delta5}
\end{gather}

Moving to the momentum representation, we can present the matrix element $\Delta V^{(1)}_2$ in the form:
\begin{gather}
\Delta\psi^{(2)}(0)=-\frac{\alpha}{2m^2}\int \tilde G_{1S}(0,r)\frac{1}{r}\hat{\bf p}^2\psi^C_{1S}(r)d{\bf r}=~~~~~~~~\\
=\psi^C_{1S}(0)\frac{-2\mu\alpha^2}{\pi m^2}\int \tilde G({\bf q},{\bf p})\frac{d{\bf q}}{(2\pi)^3}\frac{d{\bf p}}{(2\pi)^3}
\frac{d{\bf p'}}{({\bf p}-{\bf p'})^2}\left[  \frac{1}{(p'^2+W^2)}-\frac{W^2}{(p'^2+W^2)^2}  \right].~~~\nonumber
\label{delta6}
\end{gather}

Isolating the divergent term in this expression, which is determined by the first term in square brackets and the free Green's function, we calculate it taking into account the leading logarithmic term:
\begin{gather}
\Delta\psi^{(2)}(0)=\psi^C_{1S}(0)\frac{4W^2}{\pi m^2}\int^{\sim m}
\frac{d{\bf q}d{\bf p}}{(2\pi)^3(p^2+W^2)({\bf p}-{\bf p'})^2(p'^2+W^2)}=\psi^C_{1S}(0)\frac{1}{2}\alpha^2\ln\alpha^{-1} .
\label{delta7}
\end{gather}

It is useful to note that the same expression for the correction can be obtained using the relativistic correction to the wave function from \cite{lepage1}:
\begin{gather}
\Delta\psi_{rel}(k)=\psi^C(k)\frac{\alpha^2}{4}\left[  
\ln\frac{(k^2+W^2)}{W^2}+\frac{(k^2-W^2)}{kW}\arctan(\frac{k}{W})
\right].
\label{delta71}
\end{gather}

Calculating the momentum integral, we obtain in leading logarithmic order the result:
\begin{gather}
\Delta\psi_{rel}(0)=\psi^C_{1S}(0)\frac{1}{2}\alpha^2\ln\alpha^{-1}. 
\label{delta711}
\end{gather}

Other finite contributions from expression {(\ref{dv2})}, corresponding to the terms of the Coulomb Green’s function \eqref{delta2} have the form:
\begin{gather}
\Delta\psi_{rel}^{(1)}=\psi^C_{1S}(0)\left[-\frac{\alpha^2}{8}\right],
\label{delta72}
\end{gather}

\begin{gather}
\Delta\psi_{rel}^{(2)}=\psi^C_{1S}(0)\frac{W^2}{2m^2\pi^6}\int\frac{p'^2d{\bf p'}d{\bf p}d{\bf q}}{({\bf p}-{\bf p'})^2(p'^2+1)^2(q^2+1)({\bf q}-{\bf p})^2(p^2+1)}= \\
\psi^C_{1S}(0)\left[  \frac{\alpha^2}{12} (\pi^2-\frac{3}{2}) \right],  \nonumber
\label{delta73}
\end{gather}

\begin{gather}
\Delta\psi_{rel}^{(3)}=\psi^C_{1S}(0)\frac{2W^2}{m^2\pi^4}\int
R({\bf p},{\bf q})d{\bf p}d{\bf q}
\left[  \frac{2}{p}\arctan(p)-\frac{1}{p^2+1}\right] = \nonumber \\
\psi^C_{1S}(0)\left[  \frac{\alpha^2}{12} (\frac{51}{4}-\pi^2) \right]. 
\label{delta74}
\end{gather}

The second part of the operator $\Delta V_2$ also gives a finite contribution so that we represent the full contribution of this operator in the form:
\begin{gather}
\Delta\psi_2=
\psi^C_{1S}(0)\left[\frac{1}{2}\alpha^2\ln\alpha^{-1}+\frac{5}{8}\alpha^2 \right].
\label{delta75}
\end{gather}

Since we are studying the production of dimuonium and ditauonium in states with a certain spin, we also take into account in $\psi(0)$ the spin-dependent correction from the Breit Hamiltonian \eqref{dv3}.
Calculation of a matrix element with a delta function taking into account the leading logarithm (the second term in \eqref{delta2} gives the following result:
\begin{gather}
\Delta\psi^{(3)}(0)=\psi^C_{1S}(0)\left[2-\frac{7}{3}S(S+1)\right]
\frac{1}{4}\alpha^2\ln\alpha^{-1},
\label{delta8}
\end{gather}
where $S$ is the spin of the bound state which is equal to 1 for orto-leptonium.

The leading order correction in $\alpha$ is the correction for electron vacuum polarization, which is determined by the potential:
\begin{gather}
\Delta V_{vp}(r)=-\frac{\alpha^2}{3\pi}\int_1^\infty\rho(\xi)d\xi \frac{1}{r}e^{-2m_e\xi r}, ~~~
\rho(\xi)=\frac{\sqrt{\xi^2-1}(2\xi^2+1)}{\xi^4}.
\label{delta9}
\end{gather}
After analytical integration over coordinates in \eqref{psicor}, we present this correction in integral form:
\begin{gather}
\psi(0)=\psi^C_{1S}(0)\left[  a_{vp}\frac{\alpha}{\pi}  \right],~~~
r_6=\frac{m_e}{m \alpha},\\
a_{vp}=\int_1^\infty \frac{\rho(\xi)d\xi}{6(1+r_6\xi)}
\left[ 2r_6^2\xi^2+7 r_6\xi+2(1+r_6\xi)\ln(1+r_6\xi)+3 \right].
\label{delta10}
\nonumber
\end{gather}
The numerical estimate of this correction for dimuonium is determined by the electron, muon vacuum polarization, and for ditauonium by the electron, muon, $\tau$ lepton vacuum polarization in the form:
\begin{gather}
a_{vp}(e^+e^-)=0.004383,~~~a_{vp}(\mu^+\mu^-)=0.942790,~~~a_{vp}(\tau^+\tau^-)=3.613251.
\label{delta11}
\end{gather}

Summing contributions  \eqref{delta72} - \eqref{delta8} and \eqref{delta11}, we obtain a correction to the wave function at the zero of the bound state of leptons as follows:
\begin{gather}
\psi(0)=\psi^C_{1S}(0)\left\{1+a_{vp}\frac{\alpha}{\pi}+
\left(2-\frac{7}{6}S(S+1)\right)
\frac{1}{2}\alpha^2\ln\alpha^{-1}-\frac{3}{128}\alpha^2
\right\}.
\label{delta91}
\end{gather}

\section{Numerical results and Conclusions} 
\label{sec_numerical_results}

Theoretical investigation of rare exclusive decay processes of the Higgs boson represents an important task in the general direction of studying the Higgs sector in the Standard Model. In this work, the calculation of the observed Higgs boson decay widths is carried out within the framework of the relativistic quasipotential model, which we previously used in studying the pair production of mesons and baryons. The creation of bound states of particles in various reactions is also important for testing the theory of bound states in quantum field theory. The difference between the production of bound states of quarks and leptons is that in the case of production of leptons, all calculations, including various corrections, can be carried out analytically, since the wave function of the bound state is known in analytical form.

In Table~\ref{tab_decay_width} we present the numerical results of calculating the decay widths of the Higgs boson with the production of single lepton bound states and a photon or Z-boson. The obtained values for $\Gamma$ of these rare decays are small, and their observation is possible only at high-luminosity colliders being designed in the future, when the production of a significant number of Higgs bosons ($\sim 10^{10})$ will be possible \cite{clic,fcc}. Taking into account the main channel of the decay of lepton ortho-states into three photons, one can expect their reliable detection.

In Table~\ref{tab_decay_width} we also present the results of pair production of bound states of leptons, which are numerically very strongly suppressed in relation to single production processes or similar processes with the production of charmoniums or bottommoniums.
Despite the fact that there is a mass factor $1/M^2$ (M is the leptonium mass), which increases the contribution of a number of decay amplitudes (see, for example, the amplitudes in Fig.~\ref{fig4}, \ref{fig7}), nevertheless, the second square of the modulus of the wave function of the bound state contains the factor $M^4\alpha^3$ so that the pair production of lepton bound states becomes completely unlikely.

\begin{table}[htbp]
\caption{Numerical results for branching fractions}
\begin{ruledtabular}
\begin{tabular}{lcc}
Final state & Result of work \cite{higgs1} & Our result \\   \hline
$(e^{+} e^{-}) + \gamma$   &$3.5\times 10^{-12}$ & $1.10 \times 10^{-11} $\\
$(\mu^{+} \mu^{-}) + \gamma$   &$3.5\times 10^{-12} $& $1.12\times 10^{-11} $ \\
$(\tau^{+} \tau^{-}) + \gamma$  &$2.2\times 10^{-12}$ & $3.48\times 10^{-12}$ \\
$(e^{+} e^{-}) + Z$   &$5.2\times 10^{-13}$ & $7.87\times 10^{-13} $\\
$(\mu^{+} \mu^{-}) + Z$   &$5.7\times 10^{-13} $& $9.85\times 10^{-13} $ \\
$(\tau^{+} \tau^{-}) + Z$  &$1.4\times 10^{-11}$ & $5.68\times 10^{-11}$ \\
$(e^{+} e^{-}) + (e^{+} e^{-})$   & --- &  $2.05\times 10^{-19}$\\
$(\mu^{+} \mu^{-}) + (\mu^{+} \mu^{-})$ & --- & $1.13\times 10^{-20}$ \\
$(\tau^{+} \tau^{-}) + (\tau^{+} \tau^{-})$ & --- &  $ 1.09\times 10^{-18} $\\
\end{tabular}
\label{tab_decay_width}
\end{ruledtabular}
\end{table}

In this work, we investigated the role of relativistic effects caused by the law of transformation of wave functions of bound states of particles during the transition to the rest frame, relativistic corrections in the decay amplitude itself and corrections to the wave function of positronium, dimuonium and ditauonium at zero. The leading corrections are of order $\alpha$, $\alpha^2\ln\alpha$ and $\alpha^2$. In addition, along with relativistic corrections, we also took into account the effects of particle coupling in our numerical results in Table~\ref{tab_decay_width}, understanding by a purely non-relativistic calculation result one in which the mass of the bound state of leptons is equal to the sum of the lepton masses. The bound state effects occur because various mass coefficients enter into the final result in large degrees. In the case of dimuonium and ditauonium, these corrections are of order of a percent, which is mainly due to the effect of vacuum polarization of order $O(\alpha)$ in the bound state wave function. For positronium, the leading corrections are those of order $O(\alpha^2\ln\alpha,\alpha^2)$, which leads to their significant reduction to 0.01 percent.

In calculating decay widths, we examine the contribution of various decay mechanisms, the relative contribution of which changes in the transition from positronium to dimuonium and ditauonium. 
Numerous mass parameters $r_i$ in the presented analytical formulas for decay widths, which determine the final result, change greatly during the transition between lepton bound states.
But we can definitely say that the contribution from W-boson and quark loops (see Fig.~\ref{fig6},\ref{fig7}) has the greatest values, along with the contribution of the ZZ-mechanism in the case of leptonium production.
The widths of rare Higgs boson decays were calculated earlier in \cite{higgs1}.
In general, we can say that our results and \cite{higgs1} are consistent, although there are some differences in individual decay widths. These differences may be related to the different production mechanisms of lepton bound states investigated in our work.

\begin{acknowledgments}
The work is supported by the Foundation for the Advancement of Theoretical Physics and 
Mathematics “BASIS” (Grant No. 22-1-1-23-1).
The authors are grateful to D. d’Enterria and V.D. Le for useful discussions.
\end{acknowledgments}



\nocite{*}

\bibliography{pair_dilepton_production_bib}

\end{document}